\documentstyle[12pt]{article}
\begin{document}
\textheight=20 true cm
\textwidth=15 true cm
\normalbaselineskip=24 true pt
\normalbaselines
\bibliographystyle{unsrt}

% DEFINITIONS : EQUATION
\def\be {\begin{equation}}
\def\ee {\end{equation}}
\def\bea {\begin{eqnarray}}
\def\eea {\end{eqnarray}}

% DEFINITIONS : SETTING
\def\bl{\begin{flushleft}}
\def\el{\end{flushleft}}
\def\br{\begin{flushright}}
\def\er{\end{flushright}}
\def\bc{\begin{center}}
\def\ec{\end{center}}

% DEFINITIONS : SYMBOLS
\def\tg {\tilde g}
\def\tq {\tilde q}
\def\tgm {\tilde\gamma}
\def\ptg {\pi_{\tilde g}}
\def\rtg {\rho_{\tilde g}}
\def\rtgs {\rho_{\tilde g}^*}
\def\r {\rho}
\def\p {\pi}
\def\P {\Pi}
\def\G {\Gamma}
\def\a {\alpha}
\def\as {\alpha_s}

%PAPER BEGINS HERE
\setcounter{page}{0}
\thispagestyle{empty}

\br
{\sf SINP-TNP/95-04}
\er
\bc
{\bf\Large Effective Theory Approach to\\
SUSY Hadron Spectroscopy}\\[5mm]
{\large Debrupa Chakraverty, Triptesh De,\\
Binayak Dutta-Roy \footnote{Electronic address: bnyk@saha.ernet.in},
Anirban Kundu \footnote{Electronic address:
akundu@saha.ernet.in}}\\[5mm]
{\sl Theory Group, Saha Institute of Nuclear Physics,\\
1/AF Bidhannagar, Calcutta - 700 064, India}\\[5mm]
\ec

\begin{abstract}
Supersymmetric hadrons of the type $\tg\tg$, $\tg g$ and $\tg q\bar
q$ could exist depending on the masses of the gluino and the squarks
being in an appropriate range of values.
We find the energy levels of $\tg g$ and $\tg q\bar q$ bound states
(where $q$ denotes a light quark, $u$, $d$ or $s$), as well as the
strong interaction transition rates among them, using a Heavy
Gluino Effective Theory, much in the same spirit as  the well-known
Heavy Quark Effective Theory. The results are obtained with greater
ease and elegance in comparison to other approaches.
\end{abstract}
\newpage

\centerline {\large\bf 1. Introduction}

The trend of the ever-accumulating data from precision experiments
indicates that new physics beyond the Standard Model, if any,
must be of a weak and decoupling nature.
The prime candidate for such a theory is the
Supersymmetric Standard Model, either minimal or nonminimal. A huge
amount of literature \cite{haber} exists on the predicted properties and
signatures of the superpartners. However, none of them have
been experimentally detected so far, and one can only hope for
signals from the
hadronic colliders (like LHC) or $e^+e^-$ colliders (such as
LEP-200) coming into operation in the not-so-distant future.

In LHC, the most copiously produced superparticle is expected to
be the gluino,
due to its colour-octet nature and its dominant production
mechanism via strong interaction. The gluino is also regarded as
the most promising
candidate to form a supersymmetric hadron. Assuming the right- and
the left-handed squarks to be degenerate in mass, the most dominant
decay mode of the gluino is a $C$-conserving 3-body one ($\tg\rightarrow
q\bar q \tgm$). For a heavy gluino ($m_{\tg}>100$ GeV), if
the squark is also heavy (such that the factor  ${m_{\tg}^5}/
{m_{\tq}^4}\leq 100$ GeV), the decay will be suppressed by the squark
propagator and, as a result, bound states involving the
gluino ($\tg\tg$, $\tg g$, $\tg q\bar q$) can
appear in the spectra, {\em viz.}, the lifetime will be long enough
for a bound state to form. For a light gluino (1 GeV $ < m_{\tg}<4$ GeV),
the formation of SUSY hadrons would be a near certainty.

In this paper, we will analyse the nature of glueballino ($\tg g$)
and meiktino ($\tg q\bar q$) spectra using the framework of Heavy
Gluino Effective Theory (HGET) \cite{cddk2}, which, apart
from some conceptual differences outlined in the next section, is
analogous to the
well-studied Heavy Quark Effective Theory (HQET)
\cite{neubert}. The said spectra have also been analysed using the bag model
\cite{bag} and the
Bethe-Salpeter formalism  \cite{bs}; however, we will show that one
can arrive at similar, and even more predictive,
results in a  simpler and more elegant way using
an effective theory approach.

\bigskip
\centerline {\large\bf 2. SUSY hadrons}

Let us assume that the gluino undergoes only 3-body tree-level decay $\tg
\rightarrow q\bar q\tgm$. The 2-body decay mode
$\tg\rightarrow g\tgm$ will be suppressed if the right- and
left-handed squarks are degenerate in mass ($\tilde m_R=\tilde m_L$).
(For a light gluino ($m_{\tg}/m_{\tq}\ll 1$) this mode will
practically be forbidden.)
The total decay rate is given by \cite{haber}
\be
\G  (\tg\rightarrow q\bar q\tgm)={\a\as e_q^2\over 48\p}
{m_{\tg}^5\over m_{\tq}^4}\times {\cal P}
\ee
where ${\cal P}$ is the phase-space factor
\be
{\cal P}=(1-y^2)(1+2y-7y^2+20y^3-7y^4+2y^5+y^6)+24y^3(1-y+y^2)\ln y
\ee
with $y=m_{\tgm}/m_{\tg}$. We use the preferred value of
$y=1/7$. The results do not significantly
change if we vary this value, or, even if we relax the earlier
assumption that the 2-body mode is suppressed. Summing over all
possible final-state channels involving various quark-antiquark pair,
the gluino lifetime $\tau$ comes out to be
\bea
\tau &=& 1.2\times 10^{-13}{\rm s}\ \ (m_{\tg}=2~{\rm GeV},
m_{\tq}=70~{\rm GeV})\nonumber\\
\tau &=& 5.5\times 10^{-22}{\rm s}\ \ (m_{\tg}=100~{\rm GeV},
m_{\tq}=70~{\rm GeV}).
\eea
As it appears that  the light gluino window cannot  definitively be
ruled out on the basis of presently available data, we have shown one result
in that regime too, where the CDF limit on $m_{\tq}$ \cite{pdg}
slackens to 70 GeV. To form a bound state with lighter constituents,
the components have to revolve round each other at least once, and
the revolution time $\tau_{rev}$ is given by
\be
\tau_{rev}\sim 1/\as ^2\Lambda_{QCD}=1.6\times 10^{-22}{\rm s}
\ee
where we have taken $\Lambda_{QCD}=300$ MeV as a measure of the
'mass' of the light constituents. For $\tg\tg$ bound
states, $\Lambda_{QCD}$ is to be replaced by $m_{\tg}/2$, which reduces
$\tau_{rev}$ by one or more orders of magnitude, enhancing the
probability of bound state formation. Comparing (1) with (4), one
gets the condition for the formation of a bound state as
\be
m_{\tg}^5/m_{\tq}^4\leq 500~{\rm GeV}.
\ee
Taking into account the uncertainty in $\Lambda_{QCD}$, one can put a
much more conservative bound of 100 GeV on the RHS of (5). This
enables one to see that a sizable parameter space is allowed for the
existence of SUSY hadrons.

To study the spectra of the SUSY hadrons, we use the HGET framework,
developed  in a way similar to the usual HQET. Gluinos being Majorana
particles, one cannot distinguish between the particle and the
antiparticle spinor. In conventional HQET, the four-component Dirac
spinor is reduced to an effective two-component one (in the zeroth
order) by projecting out the positive energy part only. However, for
Majorana fermions, only two independent components exist,
 and hence one cannot dispense with the negative energy
part in the zeroth order of the effective Lagrangian.
 This in turn allows the fermion number violating
Green's function in the theory, both for the propagators and the vertices. In
other words, whereas in HQET  the negative-energy pole  disappears
in the limit of infinite fermion mass
(in the leading order) and there is no interaction
involving antiparticles, this is not so in HGET,  because here
one cannot differentiate between particles and antiparticles. Due to this
extra complication, some of the characteristic features of the HQET
are either
lost or modified in HGET. For example, the spin SU(2) symmetry is lost as
the relevant vertices do explicitly contain $\gamma$-matrices coming
from the charge-conjugation operator. (We recall that the spin
symmetry of HQET originates from the fact that the $q\bar q g$ vertex
involves the four-velocity $v$ of the heavy particle and
does not contain explicit $\gamma$-matrices.) This is reflected in the
calculation, for example, of the elements of the anomalous
dimension matrix; the extra diagrams as well as the extra vertex
factors contribute to the said elements.
However, the spin symmetry,  is partially
restored in the static limit $v=(1,\vec 0)$, and the elements of
the anomalous dimension matrix simplify. This
formulation will be developed in a subsequent paper \cite{cddk2},
since one does not need to be concerned
about these intricacies to study the spectra of SUSY
hadrons. The spectra, as we will show in the next section, can be
computed from a judicious comparison with the spectra of ordinary
heavy hadrons. In this paper, we consider only those  hadrons which
consist of only one heavy gluino; thus gluinonium ($\tg\tg$)
spectroscopy will not be discussed .
\bigskip
\newpage

\centerline {\large\bf 3. Sum rules}

HQET (or HGET) can be successfully applied to find the energy levels
of different hadronic states consisting of one heavy quark (gluino),
modulo the uncertainty in the off-shellness scale $\Lambda$, which is
of the order of  $\Lambda_{QCD}$. To
overcome this uncertainty, one compares the system under investigation
with an already known one, and
constructs sum rules for the masses \cite{aglietti}.

$\Lambda$ depends on the {\em brown muck}, i.e., lighter constituents
of the bound state, including sea partons. Evidently, $\Lambda$ is
different not only for 2-body glueballino and 3-body meiktino states,
but also for $1S$ and $1P$ levels of the bound states. When comparing
with the known hadrons, one should take this feature into account;
thus, $1S$ $\tg g$ states are to be compared with $1S$ $c\bar q$ or
$b\bar q$ states, while $1S$ $\tg q\bar q$ states are to be compared
with $1S$ $cq_1q_2$ (or $bq_1q_2$) states. (The same holds for
$1P$ states.)

Being a Majorana particle, the gluino has imaginary parity, and so do
meiktino and glueballino states. However, for convenience, we will
denote the parity of a state as $+(-)$ when the actual parity is
$+i(-i)$. This does not lead to any complication, since the final
decay products of $\tg$ contain a LSP (lightest supersymmetric
particle) which has imaginary intrinsic parity.

The nomenclature of states, which will be generally followed
\cite{bs} is: $1S$ $\tg u \bar d$ states
labeled as $\ptg (\rtg, \rtgs )$, where $u$ and $\bar d$ combine to
give $J_{light}=0$ $(J_{light}=1)$; $\rtg$ and $\rtgs$ denote the final
$J=1/2$ and $J=3/2$ states (fig. 2). A similar scheme defines the
$K_{\tg}$, $K_{\tg}^*$, $K_{\tg}^{**}$, $\eta_{\tg}$, $\eta'_{\tg}$,
$\omega_{\tg}$ and $\omega'_{\tg}$ states.

With the hyperfine interaction turned off, the $1S$ $\tg g$ state has
a mass $m^{(2)}_{1S}=m_{\tg}+\Lambda^{(2)}_{1S}$, where
$\Lambda^{(2)}_{1S}$, the effective $\Lambda$ for 2-body $1S$ bound
states may be estimatedfrom the known meson masses,e.g.,
\be
\Lambda^{(2)}_{1S}={1\over 4}(m_{D^0}+3m_{D^*})-m_c,
\ee
and also by
\be
\Lambda^{(2)}_{1S}={1\over 4}(m_{B^0}+3m_{B^*})-m_b,
\ee
which can be used as a cross-check. Likewise, the $1S$ meiktino
masses are given by
$m_{1S}^{(3)}=m_{\tg}+\Lambda^{(3)}_{1S}$, where similarly
\be
\Lambda^{(3)}_{1S}=m_{\Lambda_c}-m_c.
\ee
The cross-check from $\Lambda_b$ is ineffective as it has a large
uncertainty in mass \cite{pdg}.

Though the most efficient way to find $\Lambda^{(2)}_{1P}$ is from
the $D_1-D_2^*$ splitting in the case of mesons,
unfortunately, the analogous quantity for the baryons is not
accurately determinable due to the fact that $1P$ baryons
are still not very well
studied. However, it turns out that $\Lambda^{(2)}_{1P}$ is nearly
one order of magnitude smaller than $\Lambda^{(2)}_{1S}$. This is
evident from the way the $c\bar q$ mesons are split. Hyperfine
splitting due to the spin-spin interaction being a contact one,
is naturally reduced for the $1P$ states as compared to that for
the $1S$ states, and
it is the spin-orbit interaction that gives
the sizable contribution to the splitting.

As for the $1S-1P$ splitting, HQET (or HGET) cannot indicate its
magnitude; but it may be estimated from the
properties of the bound-state wavefunctions. For a linear confining
potential, it can be shown \cite{rosner} that the splitting is
constant for all hadrons of a given type (meson or baryon)
as long as one can neglect the mass of the
brown muck compared to the heavy constituent, quark or gluino. This
fact is also verified experimentally (it is well known that the
size of hadrons is nearly constant). Thus, we may take the $1S-1P$
splitting of $c\bar q$ system to be constant throughout the heavy
meson spectra, the splitting of $\Lambda_c^+(2625)-\Lambda_c^+$ to be
constant throughout the heavy baryon spectra,
and this may not be too unreasonable an assumption.

It is a phenomenological observation that the masses of the heavy
hadrons increase by about 100 MeV when one replaces one of the $u$ or
$d$ quarks by a $s$ quark, and this is explained by considering the
constituent quark masses. We assume this fact to be true for SUSY
hadrons too.

\bigskip
\centerline {\large\bf 4. Spectra of the SUSY hadrons}

First, let us assume that the gluino is massive enough (say, $m_{\tg}
= 200$ GeV) so that the hyperfine splitting is really negligible. The
spectra is shown in fig. 1. We have assumed $m_c=1.4$ GeV and
$m_b=4.7$ GeV. For $\tg g$ spectra, the positions of the $1S$ levels
are determined by $\Lambda^{(2)}_{1S}$, which is nearly 575 MeV (we
have rounded off the energy levels to a 5 MeV accuracy, which is more
than sufficient). However, $\Lambda^{(3)}_{1S}$ is 885 MeV, which is
reflected in the $I=1$ meiktino spectra. Replacing one of the first
generation quarks by a $s$ quark enhances the heavy hadron mass by
100 MeV; these are shown in the last two columns of fig. 1. The $1S-1P$
splittings are from ordinary mesonic and baryonic data.  One notes
that $I=0$ and $I=1$ $\tg u\bar d$ states, as is expected from HQET,
are degenerate in mass.

When the hyperfine interaction is turned on, the levels are split, as
shown in fig. 2. The point to note is that the $1P$ levels split very
little (as $\Lambda_{1P}$ is small), and even for $m_{\tg}=2$ GeV,
would appear as a not-too-broad band (the width being $\sim 10$ MeV). All
the $1S$ levels are fully
resolved, even to allow strong transitions among them.
Some typical strong decays (permitted by parity conservation)
among the lowest lying levels are:
$$
(1) \ptg \rightarrow (g\tg)_{{3/2}^+}+\pi;\ \ \
(2) \rtg \rightarrow (g\tg)_{{3/2}^+}+\pi;
$$
$$
(3) \rtgs \rightarrow (g\tg)_{{1/2}^+}+\pi;\ \ \
(4) (g\tg)_{{5/2}^-}\rightarrow \ptg+\pi;
$$
$$ (5) (g\tg)_{{1/2}^-}\rightarrow \rtg +\pi;\ \ \
(6) (g\tg)_{{3/2}^-}\rightarrow \rtgs +\pi;$$
$$
(7) (g\tg)_{{5/2}^-}\rightarrow {g\tg}_{{3/2}^+}+\pi\pi;\ \ \
(8) (g\tg)_{{3/2}^-}\rightarrow {g\tg}_{{1/2}^+}+\pi\pi;$$
$$ (9) (g\tg)_{{1/2}^-}\rightarrow {g\tg}_{{1/2}^+}+\pi\pi.$$
With $m_{\tg}=200$ GeV, the spin-spin and the spin-orbit splittings
are negligible, so that all of the above mentioned decays are
kinematically allowed. For a light gluino ($m_{\tg}=2$ GeV, say, as
in fig. 2), these splittings start to play significant roles, so that
transitions 2 and 6 become kinematically forbidden. However, we
stress again that the list is a typical one and by no means
exhaustive.

The decay amplitude for strong decay transitions  can be written in
HQET as \cite{neubert}
\bea
{\cal A}(s\rightarrow s'+J_h)&=& \langle\ ||\ ||\ \rangle
[(2s'+1)(2s_l+1)]^{1/2}(-)^{s_Q +s'_l+ J_h+s}\nonumber\\
&\times & \Bigg\{ {s_Q\atop J_h}\ {s'_l\atop s}\ {s'\atop s_l}\Bigg\}
\eea
where the symbols $s$, $s'$, $s_l$, $s'_l$, $s_Q$ and $J_h$ respectively
stand for total angular momentum of the initial and
the final hadron, angular momentum of the
light degrees of freedom of the initial and the
final hadron, spin of the heavy quark (=$1/2$) and total angular
momentum of the light quanta emitted; $\langle\ ||\ ||\ \rangle$
stands for the reduced matrix element. We have considered three
distinct sets of transitions: $s_l=0$ to $s'_l=0$ (1 to 3); 2-body
$s_l=1$ to $s'_l=0$ (4 to 6) and 3-body $s_l=1$ to $s'_l=0$ (7 to 9).
The reduced matrix elements are the same within a particular set but
should differ among different sets.  Without
taking into account the phase space
factors, the reduced partial widths work out to be
\bea
\G_1^r:\G_2^r:\G_3^r&=& 4:2:1\nonumber\\
\G_4^r:\G_5^r:\G_6^r&=& 1:1:0.6\\
\G_7^r:\G_8^r:\G_9^r&=& 1:0.17:1\nonumber
\eea
where the suffixes on $\G^r $ indicate the serial number of the
transitions enlisted above. However, for low $m_{\tg}$,
these results get seriously
modified due to the kinematic factor of
$(|\vec{p_\pi}|/m_{\tg})^{2J_h+1}$ which is associated with the decay
width. For large gluino mass, the kinematic factor is the same for
all members in a particular set, and so the ratios remain
unaltered. Still the $J_h=2$ decay modes (set 2) will be highly
suppressed compared to $J_h=1$ modes (set 1) due to an extra factor
of $1/m_{\tg}^2$. It is easier to show the results for 2-body decays;
for 3-body decays one can only compute the ratios at some fixed point
of the relevant Dalitz plots. For $m_{\tg}=2$ GeV, one gets
\bea
\G_1:\G_3&=& 1:8.3\nonumber\\
\G_4:\G_5&=& 2\times 10^{-4}:1
\eea
which is in sharp contrast with the n\"aive estimate of eq. (10), and
shows that some decay widths depend sensitively on $m_{\tg}$.

\bigskip
\centerline{\large\bf 5. Discussions and conclusion}

In this paper, we have obtained the energy levels and the ratios of
decay amplitudes of the SUSY hadrons, exploiting certain symmetries
of an effective-theory approach. Essentially, we have compared the
SUSY hadrons with their analogous ordinary counterparts and have
eliminated, from this comparison, the unknown factors in the
determination of the spectra. Of course, phenomenological inputs not
directly available in HQET were also used, {\em e.g.}, the fact that
$1S-1P$ splitting is nearly constant for heavy hadrons with a
linearly confining potential, or the increase of mass of the heavy
hadrons by approximately 100 MeV when a strange quark is substituted
for a lighter one ( $u$ , $d$).

The fact that the gluino is a Majorana particle hardly affects our
results, since all of them were obtained in the static limit of the
gluino where the spin symmetry is restored. However, we have not
taken into account the QCD dressing of the spectra. Most of the QCD
corrections can be taken into account with the phenomenological
choice of $\Lambda$, as the levels of ordinary hadrons, too, undergo
QCD dressing; a small part, which is really negligible, remains as
the exponents of the Wilson coefficients have some Majorana
contributions. This will be discussed in detail in our next paper
\cite{cddk2}.

According to us, the most impressive feature of an effective-theory
approach is the simplicity and elegance with which one obtains the
predictions, compared to the other approaches. The results are also
more or less in conformity with those obtained elsewhere \cite{bag,
bs}. For example, the $1S-1P$ splitting for $\tg g$ ($\tg q
\bar q$) system is 600 (300) MeV in the Bethe-Salpeter approach,
while we obtain 475 (340) MeV. The parity of the $1S$ $\tg g$ and
$1S$ $\tg q\bar q$ are same, as in the Bethe-Salpeter or the bag
model approach (however, in  the limit $m_g\rightarrow 0$, the bag
model $\tg g$ $1S$ and $1P$ turn over, changing the parity of the
ground state).

If the gluino is discovered at the future colliders, it may be
possible to study these spectra. For a heavy gluino, it may also be
possible to test how far the effective-theory approach can be
successfully extrapolated, and how large the QCD dressing effects
turn out to be. We hope that this will  provide the most severe test
for such an approach.

\newpage
\centerline{\large\bf Figure Captions}

\begin{enumerate}
\item The energy levels of SUSY hadrons for $m_{\tg}=200$ GeV,
$m_c=1.4$ GeV and $m_b=4.7$ GeV.

\item The energy levels of SUSY hadrons for $m_{\tg}=2$ GeV,
$m_c=1.4$ GeV and $m_b=4.7$ GeV.
\end{enumerate}


\newpage
\begin{thebibliography}{99}
\bibitem{haber} H.E. Haber and G.L. Kane, Phys. Rep. {\bf C117}, 75
(1985), and references therein.
\bibitem{cddk2} D. Chakraverty, T. De, B. Dutta-Roy and A. Kundu,
work in progress.
\bibitem{neubert} M. Neubert, Phys. Rep. {\bf C245}, 259 (1994);
M.B. Wise, Caltech report no. CALT-68-1901 (1993).
\bibitem{bag} F.E. Close and R.R. Horgan, Nucl. Phys. {\bf B164},
413 (1979).
\bibitem{bs} S. Ono and A.N. Mitra, Z. Phys. {\bf C25}, 245 (1984).
\bibitem{pdg} Review of Particle Properties,
Phys. Rev. {\bf D50}, 1173 (1994).
\bibitem{schuler} T. Mannel and G. Schuler, CERN report no.
CERN-TH.7468/94.
\bibitem{aglietti} U. Aglietti, Phys. Lett. {\bf B281}, 341 (1993).
\bibitem{rosner} J.L. Rosner, EFI report no. EFI 84/33 (1984)
\end{thebibliography}
\end{document}